\documentclass{article}

\usepackage{amsmath} 
\usepackage{amssymb}

\usepackage{amsfonts}
\usepackage{CJK}

\usepackage[T1]{fontenc} 
\usepackage[utf8]{inputenc} 
\usepackage{ismir,amsmath,cite,url}
\usepackage{graphicx}
\usepackage{color}

\usepackage{algorithm}
\usepackage{algorithmic}

\usepackage{lineno}
\usepackage{booktabs}

\title{SinTra: Learning an inspiration model from a single multi-track music segment}

\multauthor
{Qingwei Song$^1$ \hspace{1cm} Qiwei Sun$^2$ \hspace{1cm} Dongsheng Guo$^1$ \hspace{1cm} Haiyong Zheng$^{1*}$} 
{
$^1$ College of Electronic Engineering, Ocean University of China, China\\
$^2$ School of Microelectronics, Xi'an Jiaotong University, China\\
$^*$ Corresponding author:
{\tt\small zhenghaiyong@ouc.edu.cn}
}
\def\authorname{Q. Song, Q. Sun, D. Guo, and H. Zheng}

\begin{document}

\maketitle
\begin{abstract}
In this paper, we propose SinTra, an auto-regressive sequential generative model that can learn from a single multi-track music segment, to generate coherent, aesthetic, and variable polyphonic music of multi-instruments with an arbitrary length of bar. 
For this task, to ensure the relevance of generated samples and training music, we present a novel pitch-group representation.
SinTra, consisting of a pyramid of Transformer-XL with a multi-scale training strategy, can learn both the musical structure and the relative positional relationship between notes 
of the single training music segment. 
Additionally, for maintaining the inter-track correlation, we use the convolution operation to process multi-track music, and when decoding, the tracks are independent to each other to prevent interference. We evaluate SinTra with both subjective study and objective metrics. 
The comparison results show that our framework can learn information from a single music segment more sufficiently than Music Transformer. Also the comparison between SinTra and its variant, i.e., the single-stage SinTra with the first stage only, shows that the pyramid structure can effectively suppress overly-fragmented notes.
\end{abstract}

\begin{figure}
  \includegraphics[width=\columnwidth]{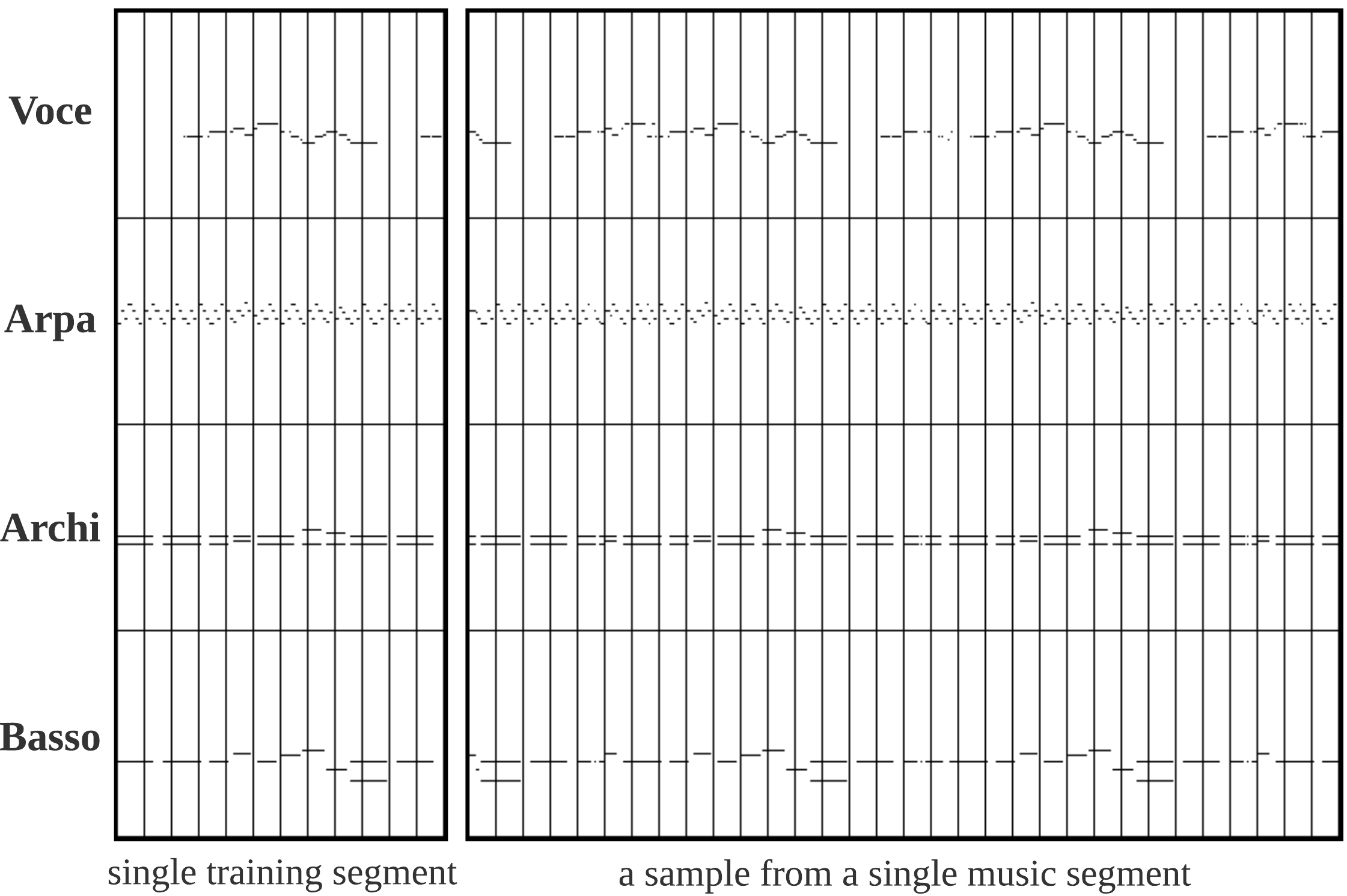}
  \caption{One sample of four-track piano-roll (right) with 32-bar length (each block represents a bar), generated by our SinTra model trained from a single music segment (left). The y-axis and x-axis represent note pitch (range from $0$ to $127$) and time step $T$, respectively.}
  \label{fig:1}
\end{figure}

\section{Introduction}\label{sec:introduction}
The current development trend of music generation is to generate harmonious multi-track music with longer-term dependency. However, in the composition of real life, the inspiration is the beginning of a song, and the composer usually creates a music based on a single music segment that comes to mind. Thus, it's more important to find ideas that are relevant to the inspiration.

As for composers, the composition process can be divided into two stages. The first stage is to generate a large number of ideas that provide inspiration for subsequent creation. The second stage is about idea convergency. Composers need to find ideas that can express their feelings and emotions to the audience from a large number of ideas, then expand, repeat and arrange them, and finally end up with a song. At the first stage, generated ideas constitute small segments of music, from which to compose more similar segments can provide more inspiration for composers. Therefore, as to music generation, it's significant to learn an inspiration model, that is, a generative model to generate music, inspiring the composers for creating a song according to a single music segment.

Recently, music generation has witnessed great progress due to the development of deep learning technologies. The mainstream methods are modeling the note sequences by drawing lessons from language models in natural language processing (NLP), which require a large number of MIDIs as training set to learn the distribution of notes. Generally, the training period is long, and the randomness of generated music is relatively high. However, if only a single music segment is available for training, it's hard for previous models to acquire enough information for leaning reasonable musical structure and relation position relationship between notes, resulting in chaotic and aesthetically unpleasant music (refer to Section~\ref{sec:results}).

At present, some one-shot generation works~\cite{shaham2019singan,awiszus2020toad} followed the multi-scale training mechanism and achieved compelling results, which allow the network to learn the information from single training data in different scales more sufficiently. 
Besides, recent works in music generation~\cite{donahue2019lakhnes,huang2020pop,wu2020transformer} adopted the Transformer-XL~\cite{dai2019transformer}, an improved variant of the Transformer~\cite{vaswani2017attention}, to introduce recurrence to the architecture, as the backbone sequence model.

In this paper, we leverage the multi-scale training scheme and the Transformer-XL architecture, to tackle the more challenging and meaningful one-shot music generation, that is, generating music from a single multi-track music segment.
To ensure the relevance of generated music and training segment, we present a novel pitch-group representation to contain all pitch group types of the single training music segment.
Besides, in order to deal with more complex multi-track music, we design three modules in each stage (scale) of our multi-scale training.

To summarize, SinTra can actually be regarded as an inspiration model for music composition. When the composer is lacking in inspiration, or in creation, it would be very repetitive to make detailed adjustments at the structure level, and SinTra can help. Certainly, the subsequent fine-tuning still needs to be done by humans. We make four contributions: (1) A novel pitch-group representation is presented to model polyphonic music of single instrument into a sequence; (2) A novel inspiration model, namely SinTra, is devised to generate meaningful music from only a single music segment; (3) Three modules based on Transformer-XL are designed for each stage of multi-scale training to process multi-track music. 
(4) The source code\footnote{\url{https://github.com/qingweisong/SinTra}} and music data are made publicly available.






\section{Related Work}
\subsection{Music Generation}
Music generation, as a niche research task of music information retrieval (MIR), has a long history and has attracted great attention in both industrial and art communities recently.

As a traditional method, Markov models are often used in the field of MIR, such as the work from Simon et al.~\cite{simon2008mysong} and Tsushima et al.~\cite{tsushima2017function}. Chuan et al.~\cite{chuan2007hybrid} have also used support vector machine (SVM) to select chord tones from given melodies.
Then, recurrent neural network (RNN)~\cite{mikolov2010recurrent} with long short-term memory (LSTM)~\cite{hochreiter1997long} and gated recurrent unit (GRU)~\cite{cho2014learning}, variational auto-encoder (VAE)~\cite{kingma2013auto}, and generative adversarial network (GAN)~\cite{goodfellow2014generative}, are common deep learning frameworks used for modeling music sequence.
MuseGAN~\cite{dong2018musegan} generated music as an image (converting MIDI into piano-roll) with GANs, and used an inter-track latent vector to make the generated multi-track music coherent.
To overcome the binarization issue, the upgraded version of MuseGAN, Binary MuseGAN~\cite{dong2018convolutional} proposed an additional refiner network, which enables generator to directly generate binary-valued piano-rolls at test time. Moreover, two types of binary neurons (BNs) considered features fewer overly-fragmented notes as compared to MuseGAN.
MIDI-Sandwich2~\cite{liang2019midi}, which also used piano-roll, applied a hierarchical multi-modal fusion generative VAE network based on RNN to collaboratively generate multi-track symbolic music.
XiaoIce Band~\cite{zhu2018xiaoice}, a melody and arrangement generation framework for pop music, introduced cooperate GRUs between each generation track to generate melody and multi-track music arrangement.
DeepJ~\cite{mao2018deepj}, based on Bi-LSTM~\cite{johnson2017generating}, was trained using piano-roll for style-specific music generating (baroque, classical, and romantic).
Different from previous music generation models, our work devotes to learning a model to generate music from only a single segment.

\begin{figure}[t]
  \includegraphics[width=\columnwidth]{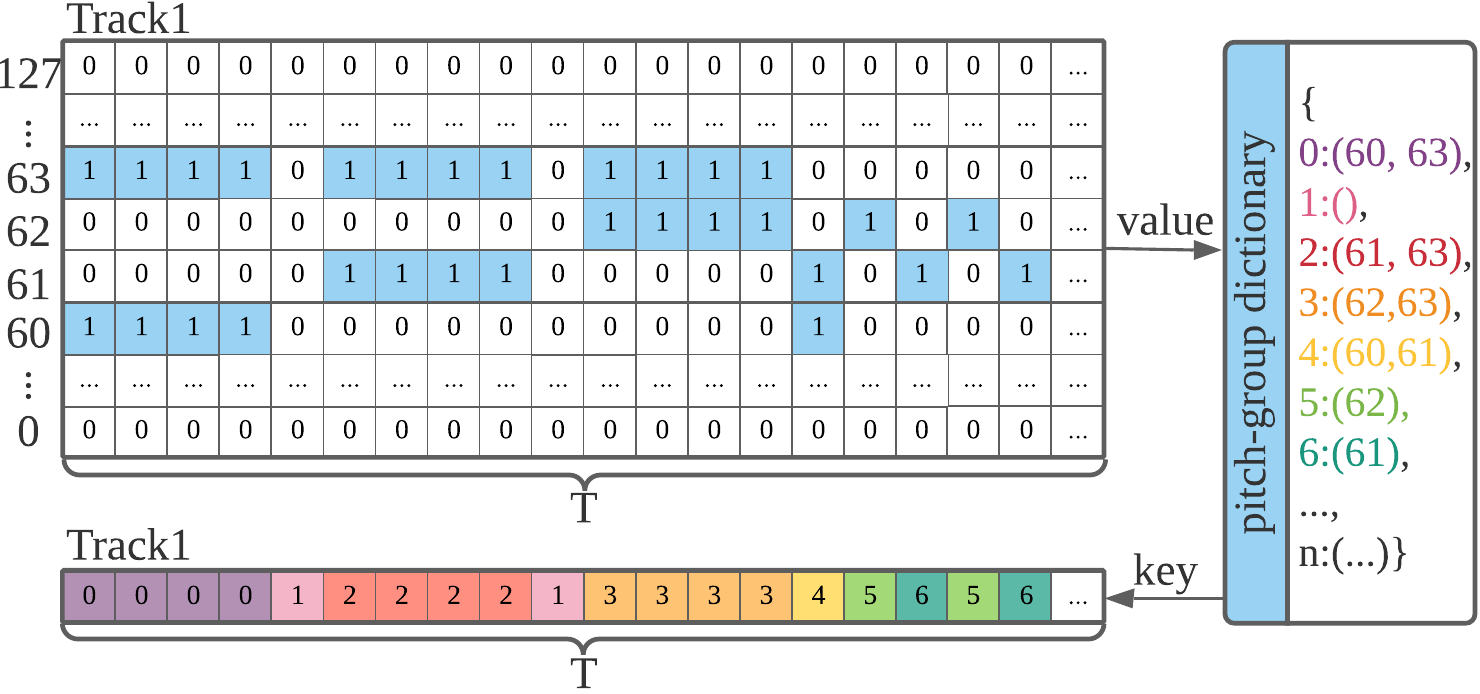}
  \caption{Illustration of our pitch-group representation. The pitch-group dictionary is built and regarded as the database of music segment containing $n+1$ key-value pairs, where the key (e.g., $0$) means the index of pitch group type and the value (e.g., $(60,63)$) means the pitch group information. Piano-roll ($T\times 128$) can be mapped to token sequence ($T\times 1$) through pitch-group dictionary ($T$ means time step). }
  \label{fig:2}
\end{figure}

\subsection{Transformer and Multi-scale Training}
Compared to LSTM or GRU, Transformer, a sequence model based on multi-head self-attention mechanism, is more parallelizable for both training and inferring, and more interpretable~\cite{vaswani2017attention}. Transformer has achieved compelling results in tasks that require maintaining long-range dependencies, such as neural machine translation~\cite{vaswani2017attention}, pre-training language models~\cite{devlin2018bert}, text-to-speech synthesis~\cite{li2019neural}, and speech recognition~\cite{mohamed2019transformers}.

For music generation, Music Transformer~\cite{huang2018music} was the first work that applies the Transformer to symbolic music generation, Huang et al. used relative positional encoding~\cite{shaw2018self} within the original Transformer architecture to capture relative timing information.
MuseNet~\cite{payne2019musenet} used sparse kernels~\cite{child2019generating} to remember the long-term structure in the composition. More recent works~\cite{donahue2019lakhnes,huang2020pop,wu2020transformer} adopted Transformer-XL~\cite{dai2019transformer} that uses recurrent memory to enable the model to attend beyond a fixed context. In this work, we also leverage the powerful long-term dependency modeling of Transformer-XL for one-shot music generation.

Notably, SinGAN~\cite{shaham2019singan} has achieved compelling results on the task of unconditional generation from a single natural image, via a pyramid of fully convolutional light-weight GANs in a coarse-to-fine fashion. 
Then TOAD-GAN~\cite{awiszus2020toad} was proposed for coherent style level generation following the one-shot training approach of SinGAN. Our work also involves the multi-scale training scheme into our SinTra model for better training from a single music segment.

\begin{figure*}[t]
  \includegraphics[width=\textwidth]{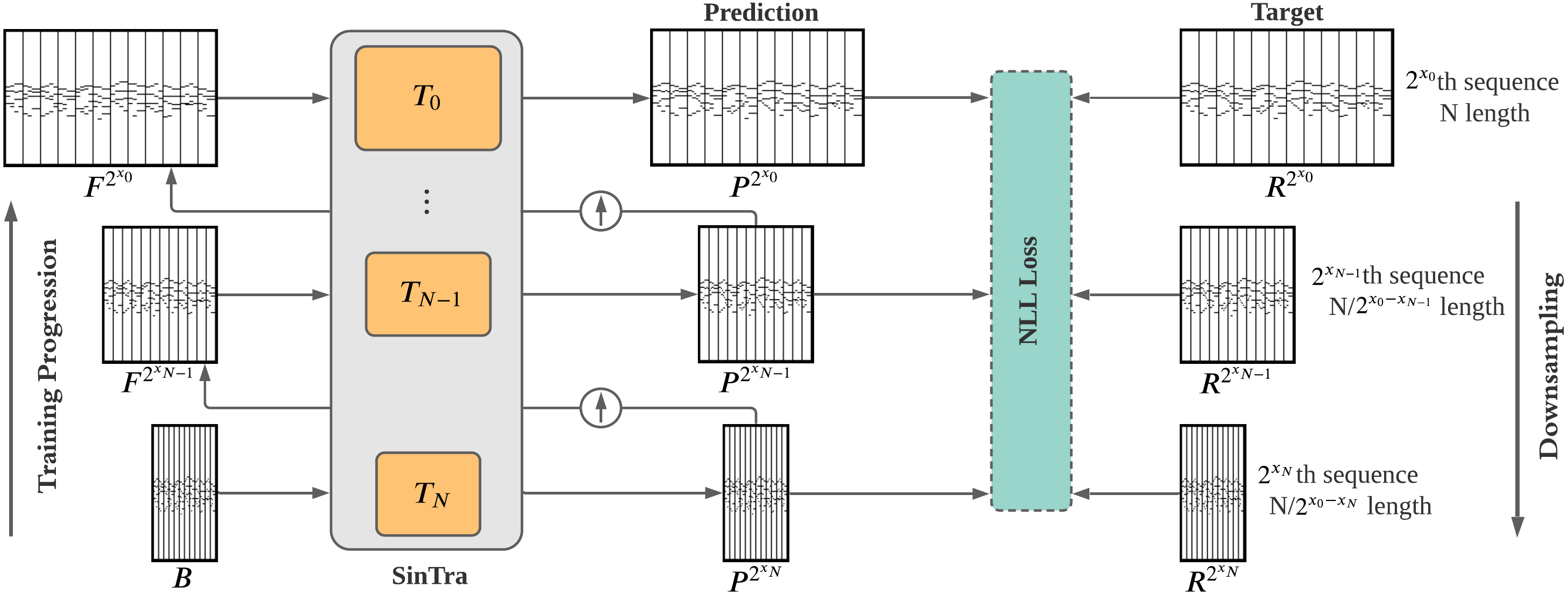}
  \caption{SinTra's multi-scale pipeline. Our model consists of a pyramid of Transformer-XLs, where both training and inference are done in a coarse-to-fine fashion. At each scale, $T_n$ learns the information of target $R^{2^{x_n}}$ by NLL loss, $2^{x_n}$th sequence means the sequence sampled by the $2^{x_n}$th note (visualize the sequence into piano-roll, $2^{x_0}>\cdots>2^{x_{N}}\geqslant4$). The input $F^{2^{x_n}}$ to $T_n$ from the previous output $P^{2^{x_{n+1}}}$, upsampled to the current temporal resolution (except for the coarsest stage which comes from B, a downsampled version of the real music). The generation process at stage $n$ involves all Transformer-XLs $\{T_N, \cdots, T_n\}$ up to this level.
  }
  \label{fig:4}
\end{figure*}

\section{Data Representation}
We use the method of language modeling to train the generation models for symbolic music. Therefore, by serializing polyphonic music into a single sequence, we express music as a series of discrete symbols determined by the data in music. 
For music generation learning from a single music segment, we present a novel pitch-group representation, which uses the index to represent various pitch group types by flexibly building a pitch-group dictionary of key-value pairs.
The principle of pitch-group representation and the construction of pitch-group dictionary are shown in \figref{fig:2}. For a segment of music, the dictionary of pitch-group representation will not be very complex, so this representation method is feasible.

We treat all types of pitch group in each time step as elements, such as \emph{harmony}, \emph{single tone} and \emph{interval}, which can use 1-dim sequence to represent polyphonic music. And the pitch-group representation can be regarded as an upgraded representation of pitch-based representation (support only monophonic music). However, this method is a double-edged sword, which will limit the output space and affect the diversity of generated music.

Piano-roll and event-based are the two most commonly data representations. Piano-roll representation used in DeepJ~\cite{mao2018deepj}, MuseGAN~\cite{dong2018musegan}, Binary MuseGAN~\cite{dong2018convolutional}, MIDI-Sandwich2~\cite{liang2019midi}, and Music Transformer~\cite{huang2018music}, is a $5$-dim matrix representation of music where the vertical and horizontal axes respectively represent note pitch and time step. 
However, the piano-roll matrix is sparse since there are many zeros, only a few notes are attacked during each time step.
Gale et al.~\cite{gale2020sparse} proved that sparse matrix has great computational potential. Treating piano-roll directly as dense matrix processing will waste computing resources.


Event-based representation used in Music Transformer~\cite{huang2018music} and LakhNES~\cite{donahue2019lakhnes}, means that the MIDI note events are converted into a sequence of tokens by a vocabulary containing 388 events. 
A one-minute song may need about 900 tokens in event-based representation. When the temporal resolution is 16th note and the tempo is 120 bpm, the piano-roll is a matrix whose shape is $(480,128)$ and the pitch-group is a sequence whose length is 480. It means that the length of the event-based is twice that of the other two methods.
Besides, although events are generated in probability order, when there is not enough training data, it's easy to generate unreasonable note (e.g., \verb|Note_on| event of the same note is generated before the \verb|Note_off| event or super long note).
In addition, \verb|TIME_SHIFT| can possibly cause the confusion of time value information as proposed by Wu et al.~\cite{wu2020transformer}.


Since traditional representation considers the universality of music representation, the coding space utilization is low when representing the information of a specific segment of music. For example, a segment of music only uses 20 pitches but still needs to use 128 pitch coding spaces.
Or only 80 events appeared, but the encoding space of 388 events still needs to be used.






\section{Music Generation learned from A Single Music Segment}\label{sec:page_size}
\subsection{Pyramid of Transformer-XL Model}

For learning reasonable musical structure and relative position relationship between pitch-group indexes of a single music segment, we adopt the multi-scale training mechanism to design a pyramid of Transformer-XLs $\{T_0, \cdots, T_N\}$. 
\figref{fig:4} shows the pipeline of SinTra for the generation of music samples. 
SinTra is trained with a sequence pyramid of each stage's real music $R$: $\{R^{2^{x_0}}, \cdots, R^{2^{x_N}}\}$ by the Negative Log Likelihood (NLL) loss, where $R^{2^{x_{n}}}$ is a downsampled version of $R^{2^ {x_0}}$. Each Transformer-XL $T_n$ is responsible of producing music samples $P^{2^n}$ with the corresponding scale of $R^{2^{x_{n}}}$.
The generation of a music sample starts at the coarsest scale and sequentially passes through all models up to the finest scale. The Transformer-XLs have the different processing length and thus capture more details as we go up the generation process. 



In order to get the sequence pyramid of $R$ for each stage, we use different note-values to sample the original sequence.
This kind of down-sampling method preserves the coarse-grained music structure information of the training music. 
We artificially define the note-value of each stage as $2^{x_n}$th, and the scale of the corresponding stage is $\frac{N}{2^{x_0-x_n}}$.
For our training, the 16th note sequence of $N$ length is sampled down to $\frac{N}{2}$ and $\frac{N}{4}$ by the 8th note and the 4th note respectively.


\subsection{Processing of Multi-track Sequences}
\begin{figure}[t]
  \includegraphics[width=\columnwidth]{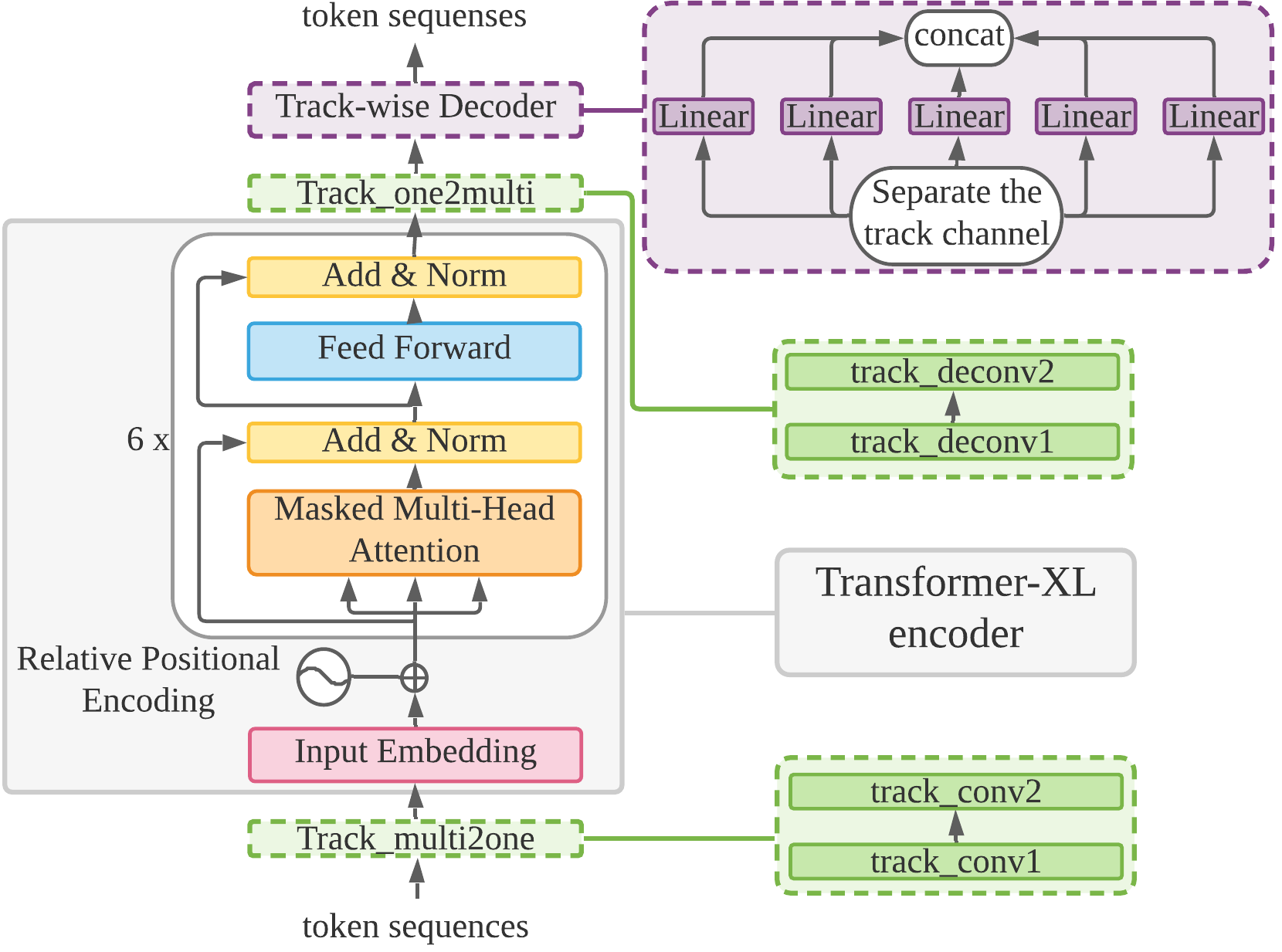}
  \caption{Network structure of each scale $T_{n}$ (left) and three modules related to multi-track sequence processing (right). After the process of \texttt{Track\_multi2one}, the shape of input sequence $(1, track, T)$ becomes to $(1, T)$. The shape of the Transformer-XL encoder output $(1, T, Feature)$ becomes to $(1,Track, T, Feature)$ after \texttt{Track\_one2multi}. Via the \texttt{Track-wise Decoder}, the final output token sequence shape is $(1, Track, T)$.} 
  \label{fig:5}
\end{figure}

All the models of each scale have a similar architecture, as depicted in \figref{fig:5}.
For maintaining the inter-track correlation, we use convolution operation to process multi-track music, and when decoding, the tracks are independent of each other to prevent interference.
The \verb|Track_multi2one| module is used to map multi-track sequence into a single sequence, and the \verb|Track_one2multi| is the inverse process of \verb|Track_multi2one|. 
The \verb|Track-wise Decoder| module is used to decode each track independently.

\subsection{Training and Inference}


The first transformer generates bar by bar sequentially and the others refine each bar in a coarse-to-fine manner.
In the $1^{st}$ scale of training, the sequence sampled by the 4th note of the $t^{th}$ bar $R^{4}_t$ is fed into the model, and the $(t+1)^{th}$ bar $R^{4}_{t+1}$ is taken as training target with NLL loss. 
In the $2^{nd}$ scale of training, to get the 8th sequence $F^{8}_{t+1}$, the output $P^{4}_{t+1}$ to the previous scale needs to be upsampled, and the model will be trained with $R^{8}_{t+1}$ as output.
In the same way, the $3^{rd}$ scale needs $F^{16}_{t+1}$ as input and to be trained with output $R^{16}_{t+1}$.
And the final output is $P^{16}_{t+1}$ from the $3^{rd}$ scale. The overall training process is shown in \eqnref{train_formula}.

\begin{equation}
\label{train_formula}
\begin{aligned} 
1st: P^{4}_{t+1}  &= Model_{1^{st}}(R^{4}_{t}) \\
     loss_{1^{st}}     &= NLL(P^{4}_{t+1}, R^{4}_{t+1}) \\
2nd: F^{8}_{t+1}  &= Upsample(P^{4}_{t+1}) \\
     P^{8}_{t+1}  &= Model_{2^{nd}}(F^{8}_{t+1}) \\
     loss_{2^{nd}}     &= NLL(P^{8}_{t+1}, R^{8}_{t+1}) \\
3rd: F^{16}_{t+1} &= Upsample(P^{8}_{t+1})    \\
     P^{16}_{t+1} &= Model_{3^{rd}}(F^{16}_{t+1}) \\
     loss_{3^{rd}}     &= NLL(P^{16}_{t+1}, R^{16}_{t+1})\\
\end{aligned}
\end{equation} 

In the inference of the NLP model, if the highest probability result is used as the prediction result every time, the generated content will repeat easily, this method is called \verb|Top|$1$. Hence, we adopt the \verb|Top|$p$ method proposed by Holtzman et al.~\cite{holtzman2019curious}, that is, the model will sample the predicted results from several of the most likely results.
In our inference process, the $1^{st}$ scale is used to predict the next bar, so we can set a larger $p=0.9$ to increase the diversity of the generated samples. However, the latter two scales are used to enhance the details, if the variety is robust, the generated content will become overly-fragmented, so we set a smaller $p=0.3$.

\section{Experiment Setups}\label{sec:typeset_text}
\subsection{Data}\label{subsec:body}
We test our method both qualitatively and quantitatively on a variety of music files in MIDI format, containing well-known works of multiple styles of music. The MIDIs that we used are taken from the JSB Chorale dataset~\cite{boulanger2012modeling}, classical music used in C-RNN-GAN~\cite{mogren2016c} from the website\footnote{\url{https://www.classicalarchives.com/}}.
We only use the JSB Chorale dataset for objective evaluation, because Music Transformer only supports single-track music. And we use all the MIDIs for subjective study.

\subsection{Model Configurations \& Training Setup}
We performed our experiments under an NVIDIA GeForce GTX TITAN X graphics card, with PyTorch 1.4.0 running under CUDA 11.2. We implemented our multi-scale training framework (\figref{fig:4}) based on the Transformer-XL encoder. The encoder has 6 layers and the number of heads is 8 with a dimension of 32. The dimension of the embedding layer is 256 and the hidden layer dimension is 1024. The dropout ratio is set at 0.09. The length of training input tokens (processing length) and the memory length of 3 stages are 4, 8, 16, respectively.

Considering downsampling, to make SinTra learn the single training segment sufficiently, it is necessary to choose the note-value of each scale reasonably.
The shorter notes appear in the training music segment, the smaller note-value used for sampling needs to be set in the finest scale, and the number of stages required for training is larger. Besides, the note-values of the adjacent scale are preferably a 2-fold relationship. Formally, the note-value of the finest scale is set to the shortest note that appears, the note-value of the $1^{st}$ stage is set to the 4th note value (standard time unit in music). 


We choose Music Transformer\footnote{Since the official code is highly coupled with Magenta, data processing and training scripts are not shown explicitly. We thus used a third-party implementation (\url{https://github.com/jason9693/MusicTransformer-pytorch}) instead.} and our variant, the model with only the first stage named single-stage SinTra, whose temporal resolution is set to the 16th note value, for comparison. 
We use Adam optimizer with $\beta_{1}=0.5$, $\beta_{2}=0.999$, $\epsilon=e^{-8}$ and follow the same learning rate schedule in Transformer-XL~\cite{dai2019transformer}. We set the number of input bars as 12 and the number of generated bars as 32.

\subsection{Subjective Study}

We set up a blind listening test for human evaluation in which test-takers listen to 7 segments of music, one from the real music, three from SinTra and three from Music Transformer. 
In the test, test-takers will be asked the same set of questions after listening to each of the two test groups, namely, to rate them on a five-point scale about the following aspects:
\begin{itemize}
\setlength{\itemsep}{0pt}
\setlength{\parsep}{0pt}
\setlength{\parskip}{0pt}
\item \textbf{Quality (Q)}: Does the generated music sound pleasing overall?

\item \textbf{Relevance (R)}: Whether the generated music give you the same feeling as the real music?

\item \textbf{Diversity (D)}: Does the generated music have new arrangement that impresses you?
\end{itemize}

Finally, we collect responses from 50 subjects, of which 20 are classified as professional composers for their musical background. The 20 professions were asked to rate each music segment they heard from the music composition theory aspect, while 30 non-composers were asked to rate their subjective feelings.

\subsection{Objective Evaluation}
Objective evaluation in music generation is still an open question, though various metrics have been proposed, the feeling of music varies from person to person, it's hard to measure the quality of generated music. 
To quantitatively compare the performance differences between the three models, we use the following metrics to measure the similarity and diversity between generated music samples and the realistic single music segment. 

\subsubsection{KL Divergence}

KL divergence measures the distance between two distributions. In this paper, pitch group indexes are used as the essential element for calculating the music distribution. The way we measure similarity is by calculating KL divergence between distributions of pitch group:
\begin{equation}
D_{kl}(P||Q)=\\\frac{1}{N_{sample}} \sum_{j=0}^{N_{sample}}\sum_{i=0}^{N_{type}}P(i)log_2(\frac{P(i)}{Q(i)}),
\label{kl_div}
\end{equation}
where $N_{sample}$ means the number of generated samples, $N_{type}$ means the number of pitch group types, $P(i)$ means the $i^{th}$ pitch group type of the generated samples, $Q(i)$ means the $i^{th}$ pitch group type of the original music.
The smaller the KL divergence, the closer the two distributions. 


\subsubsection{Pitch Group Overlap}

Referring to the idea of IoU (Intersection over Union), we design a metric named pitch group overlap:
\begin{equation}\label{overlap}
Overlap = \sum_{j=0}^{N_{}} \frac{len(set(P) \cap set(Q))}{N_{}*len(set(P) \cup set(Q))},
\end{equation}
where $set(P)$ means a set contains all pitch group types appearing in the sample, $set(Q)$ means a set contains all types appearing in the real music segment.
Overlap means the length of intersection between $set(P)$ and $set(Q)$ divided by the length of the union.
Compared to KL divergence, the overlap can measure the difference between $P$ and $Q$ distributions at a coarser granularity. The larger the overlap, the more similar the pitch group types of two songs.

It should be noted that both KL divergence and overlap can only roughly measure the similarity between the two pieces of music. In the music generation task, if the similarity is too high, the diversity will be low. Conversely, if the similarity is too low, the correlation with real music will not be ideal. Therefore, to balance diversity and relevance, it is necessary to find a suitable similarity interval.



\begin{table}[t]
 \begin{center}
 \begin{tabular}{l|l|l|l}
\toprule
               & Quality      & Relevance    & Diversity   \\
\hline
  SinTra       & 3.20/5.00    & 3.66/5.00    & 2.86/5.00   \\
  \hline
  Music Trans. & 2.34/5.00    & 2.38/5.00    & 2.54/5.00   \\
 \bottomrule
 \end{tabular}
\end{center}
\caption{Results of subjective study in a five-point scale. 
}
\label{tab:subject}
\end{table}

\section{Results and Analysis}\label{sec:results}
\subsection{Results of Subjective Study}

The results shown in \tabref{tab:subject} indicate that our SinTra receives commendable scores, especially in relevance (R). After communicating with the subjects, they reflected that there were several fluent segments in samples, but the occasional messy notes led to the decline of the overall auditory perception. 
The output space dictionary constructed in the pitch-group representation ensures the correlation between generated music and real music. Besides, the coarse-to-fine fashion introduced by the pyramid structure enables the model to fully learn the features of training data. 
Although we introduce randomness at each stage in terms of diversity (D), the results show that the generated music is still close under a single training music. 




\subsection{Comparison with Previous Work}

We compare music generation quality of SinTra with Music Transformer by: 1) we conduct experiments in the same one-shot learning condition, and since Music Transformer only supports single-track music, we use JSB Chorale dataset to evaluate the performance; 2) both models are asked to generate 10 segments in about 1 minute; 3) we set the velocity of all notes in musical pieces generated by models to a reasonable value (100). 

\begin{figure}[t]
  \includegraphics[width=\columnwidth]{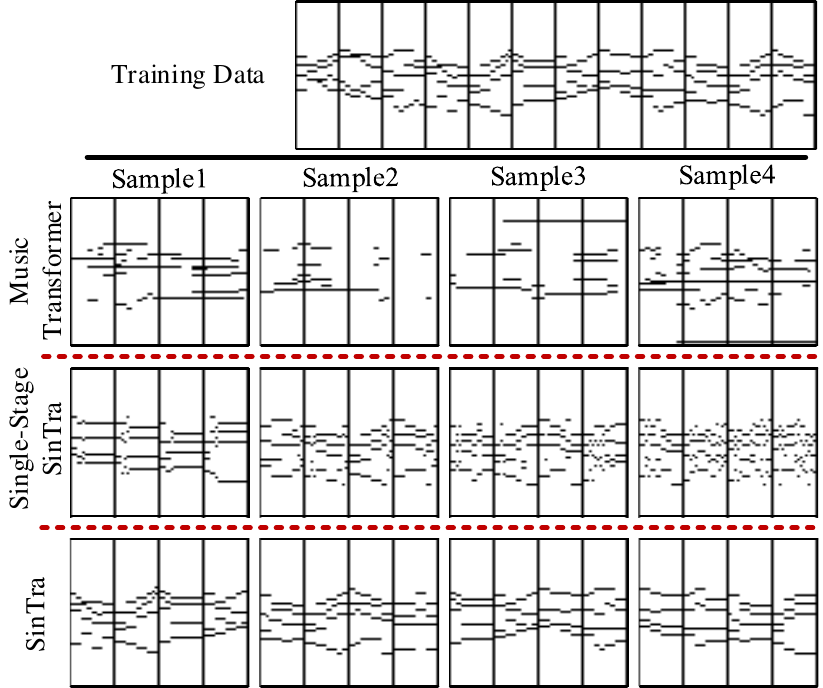}
  \caption{Samples generated by the three models. Each model lists 4 samples. All models are trained under the same training data.}
  \label{fig:samples}
\end{figure}

The comparison results are shown in \tabref{tab:metric} and the local detail of the generate music is demonstrated in \figref{fig:samples}. The convergence values of NLL loss shown in \tabref{tab:loss} also validate that Music Transformer does not fit the training music well.
The music generated by SinTra is more realistic and fits more closely with real music.

\begin{table}[t]
 \begin{center}
 \begin{tabular}{l|l|l|l}
\toprule
               & Music Trans.   & Single-stage & SinTra  \\
\hline
  KL-Div       & 40.53          & 27.84        & \textbf{21.67}   \\
  \hline
  Overlap      & 5\%           & 55\%         & \textbf{79\%}    \\
 \bottomrule
 \end{tabular}
\end{center}
 \caption{Results of objective metrics by Music Transformer, single-stage SinTra and SinTra on JSB Chorale dataset. The metrics of SinTra are marked in bold.}
 \label{tab:metric}
\end{table}

\subsection{Method Analysis}
\subsubsection{Analysis on Pyramid Structure}
To verify the effectiveness of the pyramid structure, we build the single-stage SinTra. As depicted in \figref{fig:samples}, we can see that the embryonic form of melody appears in single-stage SinTra, but is overly-fragmented and noisy. The KL divergence ($27.84$) and overlap ($55\%$) shown in \tabref{tab:metric} are not ideal because of overly-fragmented notes. The comparison results show that SinTra can effectively suppress messy notes.



\begin{table}[t]
 \begin{center}
 \begin{tabular}{l|l|l|l}
\toprule
               & Music Trans.  & Single-stage        & SinTra  \\
\hline
  NLL          & $\sim 1.0$        & $10^{-3}\sim 10^{-4}$ & $10^{-3}\sim 10^{-4}$   \\
 \bottomrule
 \end{tabular}
\end{center}
 \caption{Results of NLL loss by Music Transformer, single-stage SinTra and SinTra on JSB Chorale dataset when model converges. Both variants of SinTra can converge in $10^{-3}\sim 10^{-4}$ while Music Transformer can't.}
 \label{tab:loss}
\end{table}

\begin{figure}[t]
  \centering\includegraphics[width=0.8\columnwidth]{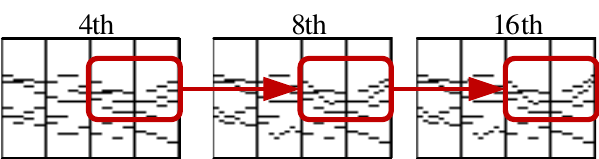}
  \caption{The effect of detail enhancement at each stage of SinTra. In the stage of the 4th note-value, simple notes will be generated but lack details, while in the 8th and 16th stages, the model refines some details.}
  \label{fig:multi_stage}
\end{figure}

\subsubsection{Analysis on Multi-stage Output}

\figref{fig:multi_stage} shows the output effect of SinTra at each stage. In the $1^{st}$ stage (the 4th note-value), the model generates simple notes, ignores local details, and sketches roughly the outline of songs. Then the $2^{nd}$ and $3^{rd}$ stages enrich the contour in turn, bringing local detail changes. This model structure can effectively solve the problem of music generated by single-stage SinTra, while ensuring the generation quality, which can introduce randomness in each stage.


\section{Conclusion and future work}
In this work, we propose SinTra, an inspiration generation framework 
to complete the task of one-shot learning in music generation. SinTra, consisting of a pyramid of Transformer-XL with a multi-scale training strategy, can learn both the musical structure and the relative positional relationship between notes of the single training music segment. 
Moreover, we present a novel pitch-group representation to ensure the relevance of generated samples and training music.
The results of subjective study and objective evaluation show the effectiveness of SinTra for learning from single training data, generating music samples with a strong correlation with the training music. 
However, there is still room for improvement in the quality and diversity of generated music.

In the future, we will study controllable music generation that can integrate emotion- and style-controlled generations into SinTra. We will also consider large-scale generative pre-training to improve generation quality. 
We hope SinTra can be leveraged to enhance musicians' productivity and inspire them to compose higher quality music.

\section{Acknowledgements}
This paper was supported by the National Natural Science Foundation of China under Grant Number 61771440.

\bibliography{ISMIRtemplate}

%
%
%
%
%

\end{document}